\documentclass[aps,prb,twocolumn,10pt,longbibliography]{revtex4-1}
\usepackage{amsmath}
\usepackage{amsfonts}
\usepackage{amssymb}
\usepackage{mathtools}
\usepackage[colorlinks=true,citecolor=blue,linkcolor=blue,urlcolor=blue]{hyperref}
\usepackage{bm}
\usepackage{times}
\usepackage[version=4]{mhchem}
\usepackage{graphicx}
\usepackage[caption=false]{subfig}
\allowdisplaybreaks%

\newcommand{\mc}{\mathcal}
\newcommand{\bra}{\langle}
\newcommand{\ket}{\rangle}
\newcommand{\ep}{\epsilon}
\newcommand{\tf}{\textbf}
\begin{document}

\title{Quantum Kinetic Theory of Nonlinear Nernst Effect}

\author{Hongchao Li}

\address{Department of Physics,University of Science and Technology of China,
China}
\email{lhcwgzg@mail.ustc.edu.cn}
\begin{abstract}
For a long period of time, we have been seeking how Berry curvature
influnces the transport properties in materials breaking time-reversal
symmetry. In time-reversal symmetric material, there will
be no thermoelectric current induced by Berry curvature in linear regime. However,
the nonlinear Hall current can be shown in non-magnetic and
non-centrosymmetric materials, where Berry curvature dipole plays
an important role. Most studies are developed from semi-classical Boltzmann equation. Here we show the quantum kinetic theory for nonlinear
Nernst effect and introduce a new type
of Berry curvature dipole: thermoelectric Berry curvature dipole. This new Berry curvature dipole will also induce the thermoelectric transport in nonlinear regime even in time-reversal invariant crystals. We will also apply our theory to topological 
crystalline insulator with tilted Dirac cone.
\end{abstract}
\maketitle
\section{Introduction}
Onsager reciprocity relation indicates that for anomalous Hall effect in linear response regime, selected material is required to break time-reversal symmetry since the Berry curvature is odd in momentum space.\cite{PhysRevResearch.2.032066}\cite{PhysRevLett.124.087402}\cite{PhysRevLett.123.246602} Therefore, the integral of Berry curvature over momentum will vanish with Fermi distribution of electrons in equilibrium. This is given by $\Omega_{a}(-k)=-\Omega_{a}(k)$. The Kramer pairs of $k$ and $-k$ are both occupied. However, according to recent research, nonlinear Hall conductivity can be still remained in time-reversal symmetric crystals. What we need is only inversion-symmetry breaking. In this case, the energy gap emerges at each Dirac node or Weyl node. More importantly, it has been found out the Berry curvature dipole is responsible for nonlinear Hall response in quantum transport by both experimental and theoretical study. \cite{PhysRevLett.115.216806}\cite{PhysRevLett.105.026805}\cite{PhysRevB.100.195117}\cite{PhysRevLett.121.246403} Indeed, there are two types
of materials creating non-trivial Berry curvature dipole. The first
kind is topological crystalline insulator SnTe, which will
undergo a ferroelectric distrotion at low temperature~\cite{PhysRevLett.122.186801},
time-reversal symmetric Weyl semimetals in the TaAs material class\cite{PhysRevLett.115.216806}
and Rashba material BiTeI\cite{PhysRevLett.121.246403} They all have strong spin-orbit
coupling contributing to their tilted Dirac cone. These tilted Dirac
cones will not change their Berry curvature but crucial to non-vanishing
dipole term. The second type is two-dimensional Dirac material without
spin-orbit coupling. Their inversion symmetry breaking attributes
to external field and substrate. Even more importantly, the appearance of a finite dipole can only be captured taking explicitly into account the terms accounting for the warping of the Fermi surface\cite{PhysRevLett.123.196403}. This new phenomena has been already studied in quantum nonlinear Hall effect\cite{PhysRevLett.115.216806} and thermal Hall effect\cite{PhysRevResearch.2.032066} with semi-classical Boltzmann equation. Inspired by the two studies, I expect to explore more on nonlinear Nernst effect. Without conventional Boltzmann equation and semi-classical approximation, I begin with generalization of the quantum kinetic theory which is more fundamental to us. With the theory in temperature field, I will investigate the nonlinear response theory in thermoelectric transport.

In this work I study the quantum kinetic theory of nonlinear Nernst
effect (NNE) in thermoelectrical transport. There is a research before on transport properties of nonlinear Nernst effect beginning from Boltzmann equation with semi-classical approximation.\cite{PhysRevB.99.201410} I will derive expression and equation of the density matrix from the basic quantum Liouville equation. I develop theory
of the nonlinear electronic transport induced by temperature gradient
in the presence of disorder. I will also introduce the new type of dipole: $\textbf{thermoelectric Berry curvature dipole}$
instead of dipole before\cite{PhysRevLett.115.216806}. This new thermoelectric Berry curvature dipole will play an important role in thermoelectric transport. This theory is also crucial to experimental physicists since they
can measure the electric current in presence of temperature gradient. I here provide a theoretical prediction of the relationship between thermoelectric conductivity and chemical potential. We can also figure out that this thermoelectric current is also connected to a term with Berry curvature which is totally different from the electric Hall effect.

This paper is organized as follows. In the second section, I will
briefly introduce not only quantum kinetic equation for Bloch electrons in the
presence of disorder, temperature gradient but also the solution
of density matrix to the equation. In the third part, I give the general expression of the density matrix by solving the quantum kinetic equation and derive the second-order response. It will also be explained that why thermoelectric Berry curvature dipole is constructed and how it influences the transport. To show the adaptability and reliability of my generalized theory, I apply the quantum kinetic theory in the presence of electric field and compare my results with that in the research before. I take disorder effect into account by applying the scattering
theory as well. I prove the terms related to Berry curvature and Berry curvature dipole have no effects on the conductivity. In the fourth section, I employ the theory before to a specific model: topological crystalline insulator, which presents
the nontrivial thermoelectric Berry curvature dipole. I will show how its thermoelectric Berry curvature dipole and thermoelectric conductivity change with chemical potential of the valley numerically. Last but not the least, I will discuss the quantum kinetic theory in more general case: non-static solution and its application to derivation of optical conductivity. This is still unfamiliar to most of the researchers since all the previous research on quantum kinetic theory only focus on the case in DC limit. Our ambition is to discover the optical current and optical conductivity in any frequency. I will also check the theory with results in semi-classical approximation.

\section{Quantum Kinetic Theory}
Without external field, the Hamiltonian
of the system is: $H=H_{0}+U$, where $U$ represents disorder potential.
The free Hamiltonian satisfies:

\begin{equation}
	H_{0}|m,\textbf{k}\rangle=\epsilon_{k}^{m}|m,\textbf{k}\rangle
\end{equation}

Here m represents band index and $\epsilon_{\textbf{k}}^{m}$ are
dispersion relationship of m-th band. In the presence of disorder,
the quantum Liouville equation after integrating the disorder's coordinates
is our beginning point~\cite{Schmidt2001}

\begin{equation}
	\frac{\partial\langle\rho\rangle}{\partial t}+\frac{i}{\hbar}[H_{0},\langle\rho\rangle]+K(\langle\rho\rangle)=0
\end{equation}

where $\langle\rho\rangle$ is density matrix after integrating all the 
disorder\cite{Schmidt2001,Liboff2003}: $\langle\rho\rangle=\frac{1}{V^{n}}\int dR_{1}...dR_{n}\rho(\textbf{r},R_{1},...,R_{n})$.
$R_{1},...,R_{n}$ are coordinates of disorder. For further step, the well-known Luttinger proposal\cite{PhysRev.135.A1505}\cite{PhysRevLett.114.196601} is introduced. To describe the thermal transport in material, I similarly add scalar potential $\psi$ which satisfies $\nabla\psi=\nabla T/T$. Therefore, the thermal field and the thermal driving term take the forms: 
\begin{equation}
	\textbf{E}_{T}=-\frac{\partial\textbf{A}_{T}}{\partial t}\equiv-\frac{\nabla T}{T}
\end{equation}
\begin{equation}
	D_{T}(\langle\rho\rangle)=\frac{1}{2\hbar}\frac{\nabla T}{T}\frac{D(\{H_{0},\langle\rho\rangle\})}{D\textbf{k}}
\end{equation}
The covariant derivative is defined as\cite{PhysRevB.96.235134}:

\begin{equation}
	\frac{DX}{D\textbf{k}}=\nabla_{\textbf{k}}X-i[\mathcal{R}_{k},X]
\end{equation}

where X is a matrix and $\mathcal{R}$ is Berry connection: $\mathcal{R}_{\textbf{k}a}^{mn}=i\langle u_{\textbf{k}}^{m}|\partial_{k_{a}}u_{\textbf{k}}^{n}\rangle$,
$\mathcal{R}_{\textbf{k}}^{mn}=\sum_{a=1}^{3}\mathcal{R}_{\textbf{k}a}^{mn}e_{a}$
Then we can construct the kinetic equation in the presence of disorder and thermal field\cite{PhysRevB.101.155204}:
\begin{equation}
	\frac{\partial\langle\rho\rangle}{\partial t}+\frac{i}{\hbar}[H_{0},\langle\rho\rangle]+K(\langle\rho\rangle)=D_{T}(\langle\rho\rangle)
\end{equation}
Here we give the numerical result of scattering term:

\begin{widetext}
	\begin{equation}
		K(\langle\rho\rangle)=\frac{1}{\hbar^{2}}\langle\int_{0}^{\infty}dt'[U,[e^{-iH_{0}t'/\hbar}Ue^{iH_{0}t'/\hbar},e^{-iH_{0}t/\hbar}\langle\rho\rangle e^{iH_{0}t/\hbar}]]\rangle
	\end{equation}

This can be decomposed into two parts:\cite{PhysRevB.96.035106}
	
	\begin{equation}
		[I(\langle\rho\rangle)]_{\textbf{k}}^{mm}=\frac{2\pi}{\hbar}\sum_{m',\textbf{k}'}\langle U_{\textbf{k}\textbf{k}'}^{mm'}U_{\textbf{k}'\textbf{k}}^{m'm}\rangle(n_{\textbf{k}}^{m}-n_{\textbf{k}'}^{m'})\delta(\epsilon_{\textbf{k}}^{m}-\epsilon_{\textbf{k}'}^{m'})
	\end{equation}
	
	\begin{equation}
		[J(\langle\rho\rangle)]_{\textbf{k}}^{mm''}=\frac{\pi}{\hbar}\sum_{m',\textbf{k}'}\langle U_{\textbf{k}\textbf{k}'}^{mm'}U_{\textbf{k}'\textbf{k}}^{m'm''}\rangle[(n_{\textbf{k}}^{m}-n_{\textbf{k}'}^{m'})\delta(\epsilon_{\textbf{k}}^{m}-\epsilon_{\textbf{k}'}^{m'})+(n_{\textbf{k}}^{m''}-n_{\textbf{k}'}^{m'})\delta(\epsilon_{\textbf{k}}^{m''}-\epsilon_{\textbf{k}'}^{m'})],~(m\neq m'')
	\end{equation}
\end{widetext}

Especially, due to energy conservation law, the main contribution is only from band-diagonal part $\langle n\rangle$. Therefore, 
the two disorder terms can be rewritten into: $I(\langle n\rangle)$ and $J(\langle n\rangle)$.
These results can also be found in \cite{Liboff2003}\cite{PhysRevB.96.235134}.

To solve the equation (6), we should sepearte the density matrix into two parts:  $\langle\rho\rangle=\langle\rho_{0}\rangle+\langle\rho_{T}\rangle$, where $\bra\rho_{0}\ket=\sum_{m}f_{0}(\xi_{\textbf{k}})|m\ket\bra m|$ represents the equilibrium-state distribution. In this passage we mainly focus on the nonequilibrium part: $\langle\rho_{T}\rangle$ induced by temperature gradient in the density matrix. The solution to this part yields:

\begin{equation}
	\langle n_{T}\rangle_{\textbf{k}}^{m}=\tau_{k}^{m}\frac{\nabla T}{T}\cdot\textbf{v}_{\textbf{k}}^{m}(\epsilon_{\textbf{k}}^{m}-\mu)\frac{\partial f_{0}(\epsilon_{k}^{m})}{\partial\epsilon_{k}^{m}}
\end{equation}
\begin{equation}
	\langle S_{T}\rangle_{\textbf{k}}^{mm'}=-i\hbar\frac{[D_{T}(\langle\rho_{0}\rangle)]_{\textbf{k}}^{mm'}-[J(\langle n_{T}\rangle)]_{\textbf{k}}^{mm'}}{\epsilon_{\textbf{k}}^{m}-\epsilon_{\textbf{k}}^{m'}}
\end{equation}

Here $\textbf{v}_{\textbf{k}}^{m}=\frac{1}{\hbar}\nabla_{\textbf{k}}\epsilon_{\textbf{k}}^{m}$,
and $\tau_{k}^{m}$ represents relaxation time which takes the form:
$1/\tau_{k}^{m}=2\pi\sum_{m'}\langle U_{\textbf{kk}'}^{mm'}U_{\textbf{k'k}}^{m'm}\rangle\int\frac{d^{d}k}{(2\pi)^{d}}\delta(\epsilon_{\textbf{k}}^{m}-\epsilon_{\textbf{k}'}^{m'})$.
This will be figured out in specific material.

Ignoring the impurity term, we can calculate the electric current
induced by temperature gradient with intrinsic velocity operator \cite{PhysRevB.101.155204}:

\[
J_{x}=\frac{1}{2}Tr[(-e)\{v_{x},\langle S_{T}\rangle\}]
\]
\begin{equation}
	=(-e)\frac{\partial_{y}T}{T}\sum_{m}\int\frac{d^{d}k}{(2\pi)^{d}}\times\Omega_{\textbf{k},z}^{m}(\epsilon_{\textbf{k}}^{m}-\mu)f_{0}(\epsilon_{\textbf{k}}^{m})
\end{equation}

Here $\{,\}$ represents anti-commutation. $\{A,B\}=AB+BA$. It is used
to prove the hermiticity of the electric current. The reason why we only pay attention to the off-diagonal part of density matrix is
no contribution comes from the diagonal part. The integral of diagonal part
is proven to be zero because it is an odd function of momentum. In addition, we can see the current is directly connected to Berry curvature of bands: $\Omega_{\textbf{k},a}^{m}=i\epsilon_{abc}\langle\partial_{k_{b}}m|\partial_{k_{c}}m\rangle$

However, this current can only be measured in time-reversal symmetry
breaking material. For time-reversal symmetric crystals, (12) will contributes
nothing. We have to consider the nonlinear Nernst effect.

\section{General theory of Nonlinear Nernst effect}

According to study by Fu \cite{PhysRevLett.115.216806},
nonlinear Hall conductivity tensor in the second harmonic term is for material preserving the time-reversal symmetry. We will here prove its reasonability with quantum kinetic theory and develop theory into thermoelectric transport

Let's focus on quantum Liouville equation first. Instead of the form
like (6), we give the general expression for the equation with temperature
gradient.

\begin{equation}
	(\mathcal{L}-D_{T})\langle\rho\rangle_{F}=D_{T}\langle\rho_{0}\rangle
\end{equation}

Here we define an operator $\mathcal{L}=P+K$, where $P\langle\rho\rangle_{F}\equiv\frac{i}{\hbar}[H_{0},\langle\rho\rangle_{F}]$.
This is accurate for $\mathcal{L}\langle\rho_{0}\rangle=0$. So we
can give the direct solution of it.

\begin{equation}
	\langle\rho\rangle_{F}=\sum_{N=1}^{\infty}(\mathcal{L}^{-1}D_{T})^{N}\langle\rho_{0}\rangle
\end{equation}

(14) is a nontrivial result for the term of $N=2$ is the response in nonlinear regime which may be related to Berry curvature dipole. The result before is just
the simplest approximation of (13). Indeed, equation (14) is obtained by iteration. In linear response theory, we just consider the $N=1$ case. Now we turn to the quadratic term.

Similarly, we can calculate the off-diagonal term(without impurity):

\begin{equation}
	\langle S_{T^{2}}\rangle=-i\hbar\frac{\partial_{y}T}{T}\sum_{nn'}\frac{\epsilon_{n'}\langle n_{T}\rangle_{\textbf{k}}^{n'}-\epsilon_{n}\langle n_{T}\rangle_{\textbf{k}}^{n}}{\epsilon_{n}-\epsilon_{n'}}\times|n\rangle\langle n|\partial_{k_{y}}n'\rangle\langle n'|
\end{equation}

Here $\epsilon_{m}=\epsilon_{k}^{m}-\mu$. In the following parts,
we note $\partial_{k_{a}}\rightarrow\partial_{a}$

We can also obtain the general expression of thermoelectric current for the quadratic term. Detailed calculation is displayed in Appendix A.
\begin{widetext}
	\begin{equation}
		J_{x}^{(2)}=\frac{1}{2}Tr[(-e){v_{x},\langle S_{T^{2}}\rangle}]=J_{x1}^{(2)}+J_{x2}^{(2)}
	\end{equation}
\begin{equation}
	J_{x1}^{(2)}=\frac{e}{2\hbar}(\frac{\partial_{y}T}{T})^{2}\sum_{m}\tau_{\textbf{k}}^{m}\epsilon_{m}^{2}f_{0}\partial_{y}\Omega_{\textbf{k}z}^{m}=e(\frac{\partial_{y}T}{T})^{2}D_{y}
\end{equation}
\begin{equation}
	J_{x2}^{(2)}=\frac{e}{\hbar}(\frac{\partial_{y}T}{T})^{2}\sum_{m}\tau_{\textbf{k}}^{m}\epsilon_{m}\partial_y\epsilon_{\tf{k}}^{m}f_{0}\Omega_{\textbf{k}z}^{m}
\end{equation}
\end{widetext}

We can see this thermoelectric currect includes two terms, which will be expanded in the following context. For simplicity, we have $\sum_{m}\int\frac{d^{d}k}{(2\pi)^{d}}\rightarrow\sum_{m}$.
Here we define the thermoelectric Berry curvature dipole: $D_{y}=\frac{1}{2\hbar}\sum_{m}\int\frac{d^{d}k}{(2\pi)^{d}}\tau_{\textbf{k}}^{m}\epsilon_{m}^{2}f_{0}\partial_{y}\Omega_{\textbf{k}z}^{m}$
. Actually, when the temperature is low enough, the contribution can be considered only from the conduction band with assumption: $\mu>0$. Since we only consider the problem on the Fermi surface, the relaxation time can be replaced with $\tau_{k_{F}}^{+}$. In this case, only the electrons near the Fermi surface on the conduction band give rise to transport. So the thermoelectric Berry curvature can be rewritten into another form.
\begin{equation}
	D_{y}=\frac{1}{2\hbar}\int\frac{d^{d}k}{(2\pi)^{d}}\epsilon_{+}^{2}f_{0}\partial_{y}\Omega_{\textbf{k}z}^{+}
\end{equation}
Here we modify the dipole by removing the relaxation time from it. We can also see another exhilirating story that this current contains a term with Berry curvature, which does not show up in the electric Hall effect. This is different from one in \cite{PhysRevLett.115.216806}
since this is thermoelectric current instead of electric current induced
by electric field. After the calculation, we still remain a question: the kinetic theory developed here should be consistent with semi-classical wavepacket dynamics\cite{PhysRevLett.95.137204}, it is crucial to check whether the result in nonlinear Hall effect is the same as that before. When the external field is electric field, we have the form of density matrix:

\begin{equation}
	\bra S_{E^2}\ket =-ieE_{y}\sum_{n,n'}\frac{\bra n_{E}\ket^{n'}_{\textbf{k}}-\bra n_{E}\ket^{n}_{\textbf{k}}}{\ep_{n}-\ep_{n'}}|n\ket\bra n|\partial_{y}n'\ket\bra n'|
\end{equation}
\[
	\bra n_{E}\ket^{m}_{\textbf{k}}=e\textbf{E}\cdot\textbf{v}_{\textbf{k}}^{m}\frac{\partial f_{0}}{\partial\textbf{k}}
\]

Since we take the zero-temperature limit, there is only contribution from conductance band. With this approximation, we obtain the same result in \cite{PhysRevLett.115.216806} by repeating the same procedure.
\[
	J_{x}=\frac{1}{2}Tr[(-e){v_{x},\bra S_{E^2}\ket}]=-\frac{1}{2\hbar}e^{3}E_{y}^{2}\sum_{n}\tau_{\textbf{k}}^{n}\frac{\partial f_{0}}{\partial k_{y}}\Omega_{\textbf{k}z}^{n}
\]
\begin{equation}
	=\frac{1}{2\hbar}e^{3}E_{y}^{2}\sum_{n}\tau_{k_{F}}^{n}\frac{\partial \Omega_{\textbf{k}z}^{n}}{\partial k_{y}}f_{0}
\end{equation}
\begin{equation}
	\chi=\frac{1}{2\hbar}e^{3}\tau_{k_{F}}^{+}\int\frac{d^{2}k}{(2\pi)^{2}}f_{0}\frac{\partial \Omega_{\textbf{k}z}^{+}}{\partial k_{y}}
\end{equation}

More details are presented in Appendix A. This also indicates that our theory is consistent with semi-classical apprximation.

The results (12) and (16-18) are derived without considering the impurity
scattering. With (11), we find the current corresponding to linear term which is induced by impurity takes the form of:
\begin{equation}
	J_{x1}^{i}=\pi e\frac{\partial_{y}T}{T}\sum_{m,m'}\bra U_{\textbf{k}\textbf{k'}}^{mm'}U_{\textbf{k'}\textbf{k}}^{m'm}\ket(n_{k}^{m}-n_{k'}^{m'})\delta(\epsilon_{\textbf{k}}^{m}-\epsilon_{\textbf{k}'}^{m'})\Omega_{kz}^{m}
\end{equation}

This is shown to be 0 after integral where $n_{\textbf{k}}^{m}=(\tau_{k_{F}}^{m})\epsilon_{m}f_{0}(\epsilon_{\textbf{k}}^{m})$.

The off-diagonal matrix elements of quadratic term induced by disorder
is given by:

\begin{equation}
	\langle S_{T^{2}}'\rangle_{k}^{mm'}=i\hbar\frac{[J(\langle n_{T^{2}}\rangle)]_{\textbf{k}}^{mm'}}{\epsilon_{\textbf{k}}^{m}-\epsilon_{\textbf{k}}^{m'}}
\end{equation}

So the current term connected to the dipole can be shown as:

\begin{equation}
	J_{x2}^{i}=\pi e(\frac{\partial_{y}T}{T})^{2}\sum_{m,m'}\bra U_{\textbf{k}\textbf{k'}}^{mm'}U_{\textbf{k'}\textbf{k}}^{m'm}\ket(N_{\textbf{k}}^{m}-N_{\textbf{k}'}^{m'})\delta(\epsilon_{\textbf{k}}^{m}-\epsilon_{\textbf{k}'}^{m'})\partial_{y}\Omega_{\textbf{k}z}^{m}
\end{equation}

Where $N_{\textbf{k}}^{m}=(\tau_{k_{F}}^{m})^{2}\epsilon_{m}^{2}f_{0}(\epsilon_{\textbf{k}}^{m})$.

This current can also be proven to be 0 after integral, which tells
that current terms in impurity scattering which is linked to dipole contributes nothing to transport. More
details are illustrated in Appendix B. We have to pay attention that here we only consider the impurity-scattering terms related to Berry curvature and dipole. They will have no effects on transport. Besides, all above is about results in DC limit. We have also developed one in the AC limit which will be discussed in the discussion section.

Since then, we have developed quantum kinetic theory of nonlinear Nernst effect
in thermoelectric transport. For further step, we are intended to
apply our results to a specific system: topological crystalline insulator.

\section{Application}

We firstly consider the Dirac semimetal materials(DSM).
In many realistic DSM, Dirac cones are more or less distorted.
The tilted Dirac cone can be realized in a number of types of materials.
To be specific, we focus on topological crystalline insulators
such as SnTe. Experiments tell us there are tilted Dirac cones on (001) surface of it. Therefore, we can calculate the thermoelectric conductance induced by
thermoelectric Berry curvature dipole.

Firstly, the low-energy model of the (001) surface is given by

\begin{equation}
	H=\xi w_{y}k_{y}\sigma_{0}+v_{x}k_{x}\sigma_{x}+\xi v_{y}k_{y}\sigma_{y}+\frac{\Delta}{2}\sigma_{z}
\end{equation}

The energy bands will take the form: $\epsilon_{\textbf{k}}^{\pm}=w_{y}k_{y}\pm\sqrt{(v_{x}k_{x})^{2}+(v_{y}k_{y})^{2}+(\frac{\Delta}{2})^{2}}$.
Here $w_{y}$ is the tilted parameter, $\Delta$ is energy gap, $v_{x},v_{y}$
represent fermi velocity in different direction. Here $\xi=\pm1$
represents the freedom of valley, which conserves time reversal(TR) symmetry
of the system. Due to the ferroelectric distortion, the Dirac
cone are turned into gapped one. Meanwhile, form of energy bands is stable since if we take influence of disorder into account, this form is still invariant. To properly account for such a dynamically generated kinetic term, we add a term $\lambda\omega\sigma_{x}$ in free fermion action. Since that, we can correct the corresponding dispersion\cite{PhysRevB.98.195123}:
\begin{equation}
	||E-tv_{y}k_{y}-\lambda E\sigma_{x}-v_{x}k_{x}\sigma_{x}-v_{y}k_{y}\sigma_{y}||=0
\end{equation}
Here we use the convention: $tv_{y}=\xi w_{y}$. After solving the equation, we have the effect energy band.
\begin{equation}
	E_{\pm}=t^{eff}v_{y}^{eff}k_{y}\pm\sqrt{(v_{x}^{eff}k_{x})^{2}+(v_{y}^{eff}k_{y})^{2}}
\end{equation}
where
\begin{equation}
	t^{eff}=\frac{t+\lambda}{1+t\lambda}
\end{equation}

\begin{equation}
	v_{y}^{eff}=\frac{1+t\lambda}{1-\lambda^{2}}v_{y}
\end{equation}

\begin{equation}
	v_{x}^{eff}=\frac{1}{\sqrt{1-\lambda^{2}}}v_{x}
\end{equation}
This method has been used by Sikkenk and Fritz to study the disorder effect in 3D tilted Weyl semimetal(WSM)\cite{PhysRevB.96.155121}. With renormalization group(RG), this term is determined to be marginal one which can not be ignored simply. However, this perturbation does nothing to the form of energy band since we can turn coefficients into effective one compared with (24).

We begin with the topological band and Berry curvature. Although we introduce the tilted parameter, the corresponding eigenvectors are still invariant:

\begin{equation}
	|\pm,\textbf{k}\rangle=\frac{1}{\sqrt{2}}\begin{pmatrix}\sqrt{1\pm\frac{\Delta}{2\epsilon_{\textbf{k}}}}\\
		\pm e^{i\theta}\sqrt{1\mp\frac{\Delta}{2\epsilon_{\textbf{k}}}}
	\end{pmatrix}
\end{equation}

The angle $\theta$ is defined by $e^{i\theta}=\frac{v_{x}k_{x}+iv_{y}k_{y}}{k_{\bot}},k_{\bot}=\sqrt{v_{x}^{2}k_{x}^{2}+v_{y}^{2}k_{y}^{2}}$.
In this way, we also have the Berry curvature which is the same one in 2D WSM.

\begin{equation}
	\Omega_{\textbf{k},z}^{\pm}=i\epsilon_{zbc}\langle\partial_{b}\pm|\partial_{c}\pm\rangle=\mp\frac{\xi\Delta v_{x}v_{y}}{4\epsilon_{\textbf{k}}}
\end{equation}

Where $\epsilon_{\textbf{k}}=\sqrt{(v_{x}k_{x})^{2}+(v_{y}k_{y})^{2}+(\frac{\Delta}{2})^{2}}$.
In general case, this will contribute nothing to transport after sum
of $\xi$. However, we will obtain the nonlinear Hall conductivity
by taking the nonlinear Hall effect into accounnt. After calculating
the relaxation time, we will give the form of conductivity. Due to
the same contribution from the different valleys, we just calculate
one and multiply it by 2. Before approaching the final result, we just make some basic assumptions: firstly, we also consider the case with low-enough temperature. Further, we assume that warping of the Fermi surface can be ignored when calculating the relaxation time.

In this way, The form of the Berry curvature dipole and conductivity are taken as:

\begin{equation}		D_{y}=\frac{1}{2}\int\frac{d^{2}k}{(2\pi)^{2}}\epsilon_{+}^{2}f_{0}(\epsilon_{\textbf{k}}^{+})\partial_{y}\Omega_{\textbf{k}z}^{+}
\end{equation}

\begin{equation}		
	\chi_1=\frac{4ev_{x}v_{y}}{n_{imp}U_{0}^{2}\mu(1+3\frac{\Delta^{2}}{4\mu^{2}})}D_{y}
\end{equation}
\begin{equation}
	\chi_2=\frac{4 ev_{x}v_{y}}{n_{imp}U_{0}^{2}\mu(1+3\frac{\Delta^{2}}{4\mu^{2}})}\int\frac{d^{d}k}{(2\pi)^{d}}\epsilon_{+}\frac{\partial\epsilon_{\textbf{k}}^+}{{\partial k_{y}}}f_{0}(\epsilon_{k}^{+})\Omega_{kz}^{+}
\end{equation}
Therefore, the total conductance is considered as: $\chi=\chi_1+\chi_2$. More details will be displayed in Appendix C. With $\partial_{y}\Omega_{\textbf{k},z}^{\pm}=\pm\frac{3\xi v_{x}v_{y}^{3}\Delta k_{y}}{4\epsilon_{\textbf{k}}^{5}}$,
we can also see that the integral will vanish if the Dirac cone is
not tilted. Although the untilted Dirac cone gives finite Berry curvature,
the Berry curvature dipole comes to zero since $\partial_{y}\Omega_{z}$
is odd function under Fermi surface.
The parameter set is $v_{x}\approx v_{y}\approx2.6328eV\cdot\mathring{A},\Delta=20meV,w_{y}=0.026328eV\cdot\mathring{A}$.

\begin{figure}[t]
	\centering
	\includegraphics[width=7.2 cm]{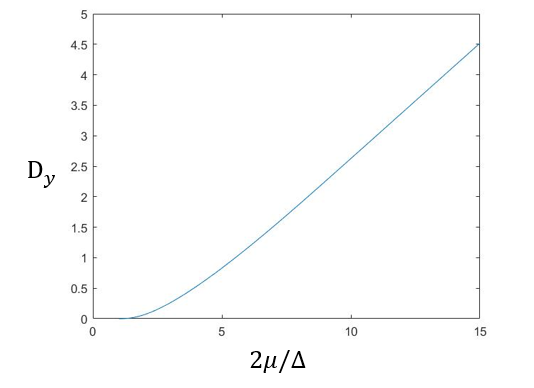}\\
	\caption{thermoelectric Berry curvature dipole(after rescaling) of SnTe as a function of $2\mu/\Delta$}
	\label{fig1}
\end{figure}

\begin{figure}[t]
	\centering
	\includegraphics[width=7.2 cm]{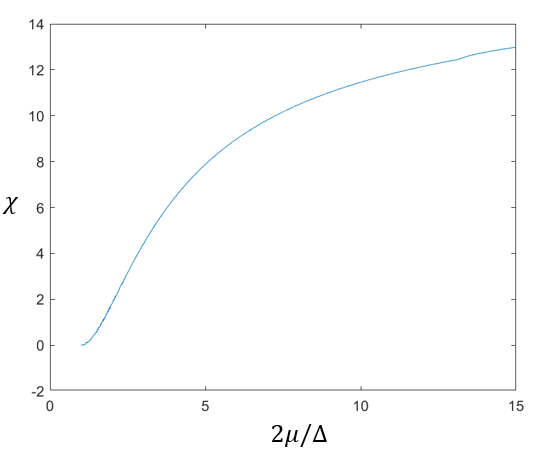}\\
	\caption{thermoelectric conductivity(after rescaling) of SnTe as a function of $2\mu/\Delta$}
	\label{fig2}
\end{figure}

\section{Discussion}

In summary, we begin with the quantum Liouville
equation and its solution in the presence of disorder and temperature
gradient. Further, we develop quantum kinetic theory of nonlinear Nernst effect with general solution to density matrix. We prove the persistence of Nernst coefficient in nonlinear regime with calculation. It also establishes a new concept of thermoelectric Berry curvature dipole dominant in the quadratic term and electric current in TR-invariant systems. The Berry curavture giving rise to linear response does not contribute to electric current without breaking time-reversal symmetry. Meanwhile, we have also proved that the main impurity scattering contributes nothing to thermoelectric transport. Finally, we apply our theory to SnTe, a topological crystalline insulator with time-reversal symmetry which has been intensively studied by recent experiments, and give numerical result of the thermoelectric Berry curvature dipole and thermoelectric conductivity. However, this theory also remains us some problems: if the external field is intensive, could we also expand the formula (13) as (14)? When the external field is in the DC limit, how could we solve the kinetic equation (6)? 

Since we only care about second-order response, we can solve equation (13) iteratively. In this way, $\bra\rho_{T}\ket=\mc{L}^{-1}D_{T}\bra\rho_{0}\ket$, $\bra\rho_{T^{2}}\ket=\mc{L}^{-1}D_{T}\bra\rho_{T}\ket=(\mc{L}^{-1}D_{T})^{2}\bra\rho_{0}\ket$, $\bra\rho_{T^{n}}\ket=\mc{L}^{-1}D_{T}\bra\rho_{T^{n-1}}\ket=(\mc{L}^{-1}D_{T})^{n}\bra\rho_{0}\ket$. In this way, we can derive any-order response iteratively. We can always derive the nonlinear response with (14) however intensive the external field is. In the AC limit, when the external field is replaced with an oscillating one $E(t)=E_{0}e^{i\omega t}$, we are still unfamiliar with the solution of the density matrix. We can solve the kinetic equation by replacing the operator $\mc{L}$ with: $\mc{M}=\mc{L}-i\omega=P+K-i\omega$ for we only consider the distribution in frequency space instead of time space. Then we can similarly derive the unsteady-state kinetic equation as: $\bra\rho\ket_{F}(\omega)=\sum_{N=1}^{\infty}(\mc{M}^{-1}D_{T})^{N}\bra\rho_{0}\ket$. However, what does the unsteady-state stand for? To further consider this problem, we firstly write down the corresponding conductivity $\sigma_{\mu\nu}(\omega)=Tr[(-e)v^{\mu}\bra\rho\ket_{F}(\omega)]/E_{0}^{\nu}$. This indicates conductivity when the external is oscillating one. In other words, this conductivity corresponds to optical conductivity in the experiments.

To clarify our quantum kinetic theory can be developed into one in AC limit, we shall calculate the second-order response in oscillating external electric field and check it with one in semi-classical approximation. We firstly ignore the scattering term to focus on the effects of $i\omega$. We firstly write down the formula for the first-order response:
\begin{equation}
	\frac{eE_{y}}{\hbar}\frac{\partial f_{0}(\epsilon_{\textbf{k}}^{m})}{\partial k_{y}}-i\omega\bra n\ket_\textbf{k}^{m}=\frac{\bra n\ket_\textbf{k}^{m}}{\tau_{k_F}^{m}}
\end{equation}
\begin{equation}
	\bra n\ket_{k}^{m}=\frac{\frac{eE_{y}}{\hbar}\frac{\partial f_{0}(\epsilon_{\textbf{k}}^{m})}{\partial k_{y}}\tau_{k_F}^{m}}{1+i\omega\tau_{k_F}^{m}}
\end{equation}
Here we figure out the relationship between the diagonal part in AC limit and that in  DC limit:$\bra n\ket_{k}^{m}(\omega)=\frac{\bra n\ket_{k}^{m}} {1+i\omega\tau_{\textbf{k}}^{m}}$. Hence we come to the conclusion that by replacing the diagonal part in (18) with generalized one, we can get the nonlinear optical conductivity. (we also assume $\mu>0$ which indicates $m=+$)
\begin{equation}
	\chi=\frac{e^{3}\tau_{k_{F}}^{+}}{2(1+i\omega\tau_{k_{F}}^{+})}\int\frac{d^{2}k}{(2\pi)^{2}}f_{0}\frac{\partial \Omega_{\textbf{k}z}^{n}}{\partial k_{y}}
\end{equation}
This is consisent with the result in \cite{PhysRevLett.115.216806}.
\section{Acknowledgement}
I acknowledge helpful discussion with Yonghao Gao at Fudan University and professor Gang Chen at Hong Kong University. My work is also supported by Physics Departement of University of Science and Technology of China. I also acknowledge my advisor Prof. Shaolong Wan in USTC and discussions on detailed calculation with other people in the group.

\begin{widetext}
	\begin{appendix}
		\section{Deriviation of (16-18)}
		
		Firstly, let's focus on the derivation of off-diagonal part of linear response$\langle S_{T}\rangle$.
		According to (14), the lowest order should be:
		
		\begin{equation}
			\langle\rho_{T}\rangle=\mathcal{L}^{-1}D_{T}(\langle\rho_{0}\rangle)
		\end{equation}
		
		Therefore, the matrix elements should be:
		
		\begin{equation}
			\langle n_{T}\rangle_{\textbf{k}}^{m}=\tau_{\textbf{k}}^{m}[D_{T}(\langle\rho_{0}\rangle)]_{\textbf{k}}^{mm}
		\end{equation}
		
		\begin{equation}
			\langle S_{T}\rangle_{\textbf{k}}^{nn'}=-i\hbar\frac{D_{T}(\langle\rho_{0}\rangle)_{\textbf{k}}^{nn'}-J(\langle n_{T}\rangle)_{\textbf{k}}^{nn'}}{\epsilon_{n}-\epsilon_{n'}}
		\end{equation}
		
		To be specific, the off-diagonal one can take another form:
		
		\begin{equation}
			\langle S_{T}\rangle=\sum_{nn'}\frac{\partial_{y}T}{T}\frac{\epsilon_{n'}f_{0n'}-\epsilon_{n}f_{0n}}{\epsilon_{n}-\epsilon_{n'}}|n\rangle\langle n|\partial_{y}n'\rangle\langle n'|
		\end{equation}
		
		Similarly, compared with (A3), we can also derive both the off-diagonal part and the diagonal part of the quadratic term.
		
		\begin{equation}
			\langle S_{T^{2}}\rangle_{\textbf{k}}^{nn'}=-i\hbar\frac{D_{T}(\langle n_{T}\rangle)_{\textbf{k}}^{nn'}-J(\langle n_{T^{2}}\rangle)_{\textbf{k}}^{nn'}}{\epsilon_{n}-\epsilon_{n'}}
		\end{equation}

		\begin{equation}
			\langle n_{T^{2}}\rangle_{\textbf{k}}^{m}=\tau_{\textbf{k}}^{m}D_{T}(\langle n_{T}\rangle)_{\textbf{k}}^{m}=\tau_{\textbf{k}}^{m}\frac{1}{2\hbar}\frac{\partial_{y}T}{T}\frac{D(\{H_{0},\langle n_{T}\rangle\})}{Dk_{y}}
		\end{equation}
		
		These are the nonlinear responses in the presence of temperature gradient. Here I remain further discussion on influence caused by impurity scattering in Appendix B. We here just care about the part induced by temperature gradient. In this way, the off-diagonal term of(A5) can be obtained as
		
		\begin{equation}
			\langle S_{T^{2}}\rangle_{\textbf{k}}^{nn'}=-i\frac{\partial_{y}T}{T}\sum_{nn'}\frac{\epsilon_{n'}\langle n_{T}\rangle_{k}^{n'}-\epsilon_{n}\langle n_{T}\rangle_{k}^{n}}{\epsilon_{n}-\epsilon_{n'}}|n\rangle\langle n|\partial_{y}n'\rangle\langle n'|
		\end{equation}
		
		So the electric current induced by thermal flow takes the form:
		
		\begin{equation}
			J_{x}^{(2)}=\frac{1}{2}Tr[(-e)\{v_{x},\langle S_{T^{2}}\rangle\}]
		\end{equation}
		
		This formula contains two parts: $\bra m|v_{x}\bra S_{T^{2}}\ket|m\ket$ and $\bra m|\bra S_{T^{2}}\ket v_{x}|m\ket$. Let's calculate the two parts separately. Since $\sum_{m}|\partial_{y}m\rangle\langle m|+|m\rangle\langle\partial_{y}m|=\partial_{y}(\sum_{m}|m\ket\bra m|)=\partial_{y}I=0$,
		we can write down another form of the intrinsic velocity operator.
		
		\begin{equation}
			v_{x}=\sum_{m'}(\epsilon_{m'}-\epsilon_{n'})|\partial_{y}m'\rangle\langle m'|+|m'\rangle\langle\partial_{y}m'|=\sum_{m'}(\epsilon_{m'}-\epsilon_{n})|\partial_{y}m'\rangle\langle m'|+|m'\rangle\langle\partial_{y}m'|
		\end{equation}
		
		\[
		\langle m|(-e)v_{x}\langle S_{T^{2}}\rangle|m\rangle=ie\frac{\partial_{y}T}{T}\sum_{m'nn'}\frac{\epsilon_{n'}\langle n_{T}\rangle_{k}^{n'}-\epsilon_{n}\langle n_{T}\rangle_{k}^{n}}{\epsilon_{n}-\epsilon_{n'}}(\epsilon_{m'}-\epsilon_{n'})\langle n|\partial_{y}n'\rangle\langle m|[|\partial_{x}m'\rangle\langle m'|+|m'\rangle\langle\partial_{x}m'|]|n\rangle\langle n'|m\rangle
		\]
		
		\[
		=-ie\frac{\partial_{y}T}{T}\sum_{m'nn'}[\epsilon_{n'}\langle n_{T}\rangle_{k}^{n'}-\epsilon_{n}\langle n_{T}\rangle_{k}^{n}]\langle n|\partial_{y}m\rangle\langle m|\partial_{x}n\rangle\delta_{m'n}\delta_{n'm}
		\]
		
		\[
		=ie\frac{\partial_{y}T}{T}\sum_{n}[\epsilon_{n'}\langle n_{T}\rangle_{k}^{n'}-\epsilon_{n}\langle n_{T}\rangle_{k}^{n}]\langle\partial_{x}m|n\rangle\langle n|\partial_{y}m\rangle
		\]
		
		\begin{equation}
			=-ie\frac{\partial_{y}T}{T}[\epsilon_{m}\langle n_{T}\rangle_{k}^{m}\langle\partial_{x}m|\partial_{y}m\rangle-\sum_{n}\epsilon_{n}\langle n_{T}\rangle_{k}^{n}\langle\partial_{x}m|n\rangle\langle n|\partial_{y}m\rangle]
		\end{equation}
		
		\[
		\langle m|(-e)\langle S_{T^{2}}\rangle v_{x}|m\rangle=ie\frac{\partial_{y}T}{T}\sum_{m'nn'}\frac{\epsilon_{n'}\langle n_{T}\rangle_{k}^{n'}-\epsilon_{n}\langle n_{T}\rangle_{k}^{n}}{\epsilon_{n}-\epsilon_{n'}}(\epsilon_{m'}-\epsilon_{n})\langle\partial_{y}n|n'\rangle\langle m|n\rangle\langle n'|[|\partial_{x}m'\rangle\langle m'|+|m'\rangle\langle\partial_{x}m'|]|m\rangle
		\]
		
		\begin{equation}
			=ie\frac{\partial_{y}T}{T}[\epsilon_{m}\langle n_{T}\rangle_{k}^{m}\langle\partial_{y}m|\partial_{x}m\rangle-\sum_{n}\epsilon_{n}\langle n_{T}\rangle_{k}^{n}\langle\partial_{y}m|n\rangle\langle n|\partial_{x}m\rangle]
		\end{equation}
		
		The first term of (A10,A11) can be pointed out as Berry curvature:
		
		\begin{equation}
			-ie\frac{\partial_{y}T}{T}\sum_{m}[\epsilon_{m}\langle n_{T}\rangle_{k}^{m}\langle\partial_{x}m|\partial_{y}m\rangle-\epsilon_{m}\langle n_{T}\rangle_{k}^{m}\langle\partial_{y}m|\partial_{x}m\rangle]=-e\frac{\partial_{y}T}{T}\sum_{m}\epsilon_{m}\langle n_{T}\rangle_{k}^{m}\Omega_{\textbf{k},z}^{m}
		\end{equation}
		
		Due to the sum of index m, the second term of (A10,A11) can be combined
		together.
		
		\[
		ie\frac{\partial_{y}T}{T}\sum_{n,m}(\epsilon_{n}\langle n_{T}\rangle_{k}^{n}\langle\partial_{x}m|n\rangle\langle n|\partial_{y}m\rangle-\epsilon_{n}\langle n_{T}\rangle_{k}^{n}\langle\partial_{y}m|n\rangle\langle n|\partial_{x}m\rangle)
		\]
		
		\[
		=ie\frac{\partial_{y}T}{T}\sum_{n,m}(\epsilon_{n}\langle n_{T}\rangle_{k}^{n}\langle\partial_{y}n|m\rangle\langle m|\partial_{x}n\rangle-\epsilon_{n}\langle n_{T}\rangle_{k}^{n}\langle\partial_{x}n|m\rangle\langle m|\partial_{y}n\rangle)
		\]
		
		\[
		=ie\frac{\partial_{y}T}{T}\sum_{n}(\epsilon_{n}\langle n_{T}\rangle_{k}^{n}\langle\partial_{y}n|\partial_{x}n\rangle-\epsilon_{n}\langle n_{T}\rangle_{k}^{n}\langle\partial_{x}n|\partial_{y}n\rangle)
		\]
		
		\begin{equation}
			=-e\frac{\partial_{y}T}{T}\sum_{n}\epsilon_{n}\langle n_{T}\rangle_{k}^{n}\Omega_{\textbf{k},z}^{m}
		\end{equation}
		
		Correspondingly, the current can be figured out:
		
		\begin{equation}
			J_{x}^{(2)}=-\frac{e}{2}(\frac{\partial_{y}T}{T})^{2}\sum_{m}\int\frac{d^{d}k}{(2\pi)^{d}}\epsilon_{m}^{2}\tau_{k}^{m}\frac{\partial f_{0}(\epsilon_{k}^{m})}{\partial k_{y}}\Omega_{kz}^{m}
		\end{equation}
		
		where $\Omega_{\textbf{k}z}^{m}=i(\langle\partial_{x}m|\partial_{y}m\rangle-\langle\partial_{y}m|\partial_{x}m\rangle)$
		represents Berry curvature.
		This can be separated into two parts if we basically assume only the conductance band contributes:
		\begin{equation}
			J_{x1}^{(2)}=\frac{e}{2}(\frac{\partial_{y}T}{T})^{2}\tau_{k_F}^{+}\sum_{m}\int\frac{d^{d}k}{(2\pi)^{d}}\epsilon_{m}^{2} f_{0}(\epsilon_{k}^{m})\frac{\partial\Omega_{kz}^{m}}{{\partial k_{y}}}
		\end{equation}
		\begin{equation}
		J_{x2}^{(2)}=e(\frac{\partial_{y}T}{T})^{2}\tau_{k_F}^{+}\sum_{m}\int\frac{d^{d}k}{(2\pi)^{d}}\epsilon_{m}\frac{\partial\epsilon_{\textbf{k}}^m}{{\partial k_{y}}} f_{0}(\epsilon_{k}^{m})\partial\Omega_{kz}^{m}
		\end{equation}
		In this way, we have the form of thermoelectric conductivity with (A15, A16)
		
		\begin{equation}
			\chi_1=\frac{e}{2}\tau_{k_{F}}^{+}\int\frac{d^{d}k}{(2\pi)^{d}}\epsilon_{+}^{2}f_{0}(\epsilon_{\textbf{k}}^{+})\partial_{y}\Omega_{\textbf{k},z}^{+}=e\tau_{k_{F}}^{+}D_{y}
		\end{equation}
		\begin{equation}
			\chi_2=e\tau_{k_F}^{+}\sum_{m}\int\frac{d^{d}k}{(2\pi)^{d}}\epsilon_{m}\frac{\partial\epsilon_{\textbf{k}}^m}{{\partial k_{y}}}f_{0}(\epsilon_{k}^{m})\Omega_{kz}^{m}
		\end{equation}
		Here $D_{y}=\frac{1}{2}\int\frac{d^{d}k}{(2\pi)^{d}}\epsilon_{+}^{2}f_{0}(\epsilon_{\textbf{k}}^{+})\partial_{y}\Omega_{\textbf{k},z}^{+}$
		represents the new type of Berry curvature dipole. We can see there are two terms in the thermoelectric conductivity. The first one is connected to the thermoelectric Berry curvature dipole. The second one, which does not show up in electric Hall effect, is directly linked to the Berry curvature.
		
		\section{Derivation of (23,25)}
		
		At the beginning, let's focus on the impurity scattering of the linear
		term.
		
		\begin{equation}
			J_{x1}^{i}=\frac{1}{2}Tr[(-e){v_{x},\langle S_{T}'\rangle}]
		\end{equation}
		
		\begin{equation}
			\langle S_{T}'\rangle=i\hbar\sum_{nn'}\frac{[J(\langle n_{T}\rangle)]_{\textbf{k}}^{nn'}}{\epsilon_{n}-\epsilon_{n'}}|n\rangle\langle n'|
		\end{equation}
		
		With intrinsic velocity(A9), you can also obtain the current density(B1). To begin with, let's firstly consider the related two terms: $\langle m|v_{x}\langle S_{T}'\rangle|m\rangle$ and $\langle m|\langle S_{T}'\rangle v_{x}|m\rangle$:
		
		\[
		\langle m|v_{x}\langle S_{T}'\rangle|m\rangle=i\sum_{m'nn'}(\epsilon_{m'}-\epsilon_{n'})\frac{[J(\langle n_{T}\rangle)]_{\textbf{k}}^{nn'}}{\epsilon_{n}-\epsilon_{n'}}\langle m|[|\partial_{x}m'\rangle\langle m'|+|m'\rangle\langle\partial_{x}m'|]|n\rangle\langle n'|m\rangle
		\]
		
		\begin{equation}
			=i\sum_{n}[J(\langle n_{T}\rangle)]_{\textbf{k}}^{nm}\langle m|\partial_{x}n\rangle
		\end{equation}
		
		\[
		\langle m|\langle S_{T}'\rangle v_{x}|m\rangle=i\sum_{m'nn'}(\epsilon_{m'}-\epsilon_{n})\frac{[J(\langle n_{T}\rangle)]_{\textbf{k}}^{nn'}}{\epsilon_{n}-\epsilon_{n'}}\langle m|n\rangle\langle n'|[|\partial_{x}m'\rangle\langle m'|+|m'\rangle\langle\partial_{x}m'|]|m\rangle
		\]
		
		\begin{equation}
			=-i\sum_{n}[J(\langle n_{T}\rangle)]_{\textbf{k}}^{nm}\langle\partial_{x}n|m\rangle
		\end{equation}
		
		With (B3,B4) we can directly figure out the form of .
		
		\begin{equation}
			J_{x1}^{i}=e\sum_{n,m}\int\frac{d^{d}k}{(2\pi)^{d}}[J(\langle n_{T}\rangle)]_{\textbf{k}}^{nm}Im(\langle m|\partial_{x}n\rangle)
		\end{equation}
		
		If we take the zero-temperature approximation, the non-equilibrium
		distribution induced by temperature gradient will just take place
		on the conductance band. So this will contribute nothing after sum
		of band index. However, if we take the band-diagonal part into account,
		we can find the term related to Berry curvature.We assume the Born
		approximation: $\bra U(\textbf{r})U(\textbf{r'})\ket=n_{imp}U_0^{2}\delta(\textbf{r}-\textbf{r'})$\cite{PhysRevB.97.201301}
		Then we have:
		\begin{equation}
			\bra U_{\textbf{k}\textbf{k'}}^{mm'}U_{\textbf{k'}\textbf{k}}^{m'm}\ket=n_{imp}U_{0}^{2}\bra u_{\textbf{k}}^{m}|u_{\textbf{k'}}^{m'}\ket\bra u_{\textbf{k'}}^{m'}|u_{\textbf{k}}^{m}\ket
		\end{equation}
	In most cases, this is not a trivial result since we may have strong SOC in the material. We can not simply come to the general case. However, we can still make some approximation. Firstly, the temperature is low enough that we can still replace the relaxation time with one on the Fermi surface which is noted by $\tau_{k_{F}}^{m}$. Further, even though the Fermi surface is partly distorted, we still assume the diagonal part $\bra n_{T}\ket_{\textbf{k}}^{m}$ and $\bra n_{T^{2}}\ket_{\textbf{k}}^{m}$ are approximately considered as functions of $\epsilon_{\textbf{k}}^{m}$. 
		\[
		J_{x1}^{i}=-i\pi e\sum_{m,m'}\langle U_{\textbf{k}\textbf{k}'}^{mm'}U_{\textbf{k}'\textbf{k}}^{m'm}\rangle(\langle n_{T}\rangle_{\textbf{k}}^{m}-\langle n_{T}\rangle_{\textbf{k}'}^{m'})\delta(\epsilon_{\textbf{k}}^{m}-\epsilon_{\textbf{k}'}^{m'})(\langle m|\partial_{x}m'\rangle-\langle\partial_{x}m'|m\rangle)
		\]
		
		\begin{equation}
			=-i\pi e\sum_{m}\langle U_{\textbf{k}\textbf{k}'}^{mm}U_{\textbf{k}'\textbf{k}}^{mm}\rangle(\langle n_{T}\rangle_{\textbf{k}}^{m}-\langle n_{T}\rangle_{\textbf{k}'}^{m})\delta(\epsilon_{\textbf{k}}^{m}-\epsilon_{\textbf{k}'}^{m})(\langle m|\partial_{x}m\rangle-\langle\partial_{x}m|m\rangle)
		\end{equation}
		
		\begin{equation}
			\rightarrow\pi e\frac{\partial_{y}T}{T}\sum_{m}\tau_{k_{F}}^{m}(n_{\textbf{k}}^{m}-n_{\tf{k'}}^{m'})\delta(\epsilon_{\textbf{k}}^{m}-\epsilon_{\textbf{k}'}^{m})\Omega_{\tf{k},z}^{m}
		\end{equation}
		
		where $n_{\textbf{k}}^{m}=(\tau_{k_{F}}^{m})\epsilon_{m}f_{0}(\epsilon_{\textbf{k}}^{m})$,
		This is exactly the equation(21), which is only connected to $\epsilon_{\textbf{k}}^{m}$. Therefore,
		this current will vanish after integrating $\textbf{k'}$.
		
		We can clearly see that the term connected with Berry curvature has no effect on transport in linear regime. However, current from other terms is still unknown to us in general case. Hence, we can only consider other terms in specific models.
		
		Let's turn to the band-diagonal part of the quadratic term.
		
		\begin{equation}
			\langle n_{T^{2}}\rangle_{\textbf{k}}^{m}=\frac{\tau_{\textbf{k}}^{m}}{\hbar}\frac{\partial_{y}T}{T}[\frac{D(H_{0}\langle n_{T}\rangle)}{Dk_{y}}]_{\textbf{k}}^{m}=\frac{(\tau_{k_{F}}^{m})^{2}}{\hbar}(\frac{\partial_{y}T}{T})^{2}\partial_{y}(\epsilon_{m}^{2}\partial_{y}f_{0}(\epsilon_{\textbf{k}}^{m}))
		\end{equation}
		
		In this way, the impurity scattering part is derived as:
		
		\begin{equation}
			\langle S_{T^{2}}'\rangle_{k}^{mm'}=i\hbar\frac{[J(\langle n_{T^{2}}\rangle)]_{\textbf{k}}^{mm'}}{\epsilon_{\textbf{k}}^{m}-\epsilon_{\textbf{k}}^{m'}}
		\end{equation}
		
		\begin{equation}
			\langle S_{T^{2}}'\rangle=i\pi\sum_{mm'm''}\sum_{\textbf{k}'}\langle U_{\textbf{kk'}}^{mm'}U_{\textbf{k'k}}^{m'm}\rangle\frac{(n_{\textbf{k}}^{(2)m}-n_{\textbf{k}'}^{(2)m'})\delta(\epsilon_{\textbf{k}}^{m}-\epsilon_{\textbf{k}'}^{m'})+(n_{\textbf{k}}^{(2)m''}-n_{\textbf{k}'}^{(2)m'})\delta(\epsilon_{\textbf{k}}^{m''}-\epsilon_{\textbf{k}'}^{m'})}{\epsilon_{m}-\epsilon_{m''}}(|\partial_{y}m\rangle\langle m''|+|m\rangle\langle\partial_{y}m''|)
		\end{equation}
		
		Here $n_{\textbf{k}}^{(2)m}$ is not the one in quadratic term. It is
		given by:
		
		\begin{equation}
			n_{\textbf{k}}^{(2)m}=\frac{(\tau_{k_{F}}^{m})^{2}}{\hbar}(\frac{\partial_{y}T}{T})^{2}\epsilon_{m}^{2}\partial_{y}f_{0}(\epsilon_{\textbf{k}}^{m})
		\end{equation}
		
		So we can calculate each matrix element of it.(Here we also use the
		simpliest approximation)
		
		\begin{equation}
			\langle S_{T^{2}}'\rangle_{\textbf{k}}^{nn'}=i\pi \sum_{m,m',m''}\sum_{\textbf{k'}}\langle U_{\textbf{kk'}}^{mm'}U_{\textbf{k'k}}^{m'm}\rangle[\frac{g(m,m',n')}{\epsilon_{m}-\epsilon_{n'}}\langle n|\partial_{y}m\rangle\delta_{m''n'}+\frac{g(n,m',m'')}{\epsilon_{n}-\epsilon_{m''}}\langle\partial_{y}m''|n'\rangle\delta_{mn}]
		\end{equation}
		
		With $g(m,m',n')=(n_{k}^{(2)m}-n_{k'}^{(2)m'})\delta(\epsilon_{k}^{m}-\epsilon_{k'}^{m'})+(n_{k}^{(2)m''}-n_{k'}^{(2)m'})\delta(\epsilon_{k}^{m''}-\epsilon_{k'}^{m'})$.
		The intrinsic contribution to velocity operator in the eigenstate
		basis is
		
		\begin{equation}
			v_{x}=\sum_{l'}(\epsilon_{l'}-\epsilon_{m''})[|\partial_{x}l'\rangle\langle l'|+|l'\rangle\langle\partial_{x}l'|]
		\end{equation}
		
		So the diagonal part is given by
		
		\[
		\langle l|v_{x}\langle S_{T^{2}}'\rangle|l\rangle=i\pi \sum_{l'n'n}\sum_{m,m',m''}\langle U_{\textbf{kk'}}^{mm'}U_{\textbf{k'k}}^{m'm}\rangle[\frac{g(m,m',n')}{\epsilon_{m}-\epsilon_{n'}}\langle n|\partial_{y}m\rangle\delta_{m''n'}+\frac{g(n,m',m'')}{\epsilon_{n}-\epsilon_{m''}}\langle\partial_{y}m''|n'\rangle\delta_{mn}]
		\]
		
		\begin{equation}
			\times(\epsilon_{l'}-\epsilon_{m''})\langle l|[|\partial_{x}l'\rangle\langle l'|+|l'\rangle\langle\partial_{x}l'|][|n\rangle\langle n'|\|l\rangle
		\end{equation}
		
		With tedious calculation, we obtain the only term seemingly connected
		to Berry curvature dipole.
		
		\begin{equation}
			i\pi\sum_{l,m,m',m''}\langle U_{\textbf{kk'}}^{mm'}U_{\textbf{k'k}}^{m'm}\rangle g(m,m',m'')\langle\partial_{y}m''|l\rangle\langle l|\partial_{x}m\rangle
		\end{equation}
		
		In this way, the current related to impurity scattering can be figured
		out:
		
		\begin{equation}
			J_{x2}^{i}=\frac{1}{2}Tr[(-e)\{v_{x},\langle S_{T^{2}}'\rangle\}]=\pi e\sum_{m,m'}\langle U_{\textbf{kk'}}^{mm'}U_{\textbf{k'k}}^{m'm}\rangle(N_{\textbf{k}}^{m}-N_{\textbf{k'}}^{m'})\delta(\epsilon_{\textbf{k}}^{m}-\epsilon_{\textbf{k'}}^{m'})\partial_{y}\Omega_{\textbf{k},z}
		\end{equation}
		
		where $N_{\textbf{k}}^{m}=(\tau_{k_{F}}^{m})^{2}\epsilon_{m}^{2}f_{0}(\epsilon_{\textbf{k}}^{m})$,
		which is only connected with $\epsilon_{\textbf{k}}^{m}$. Therefore,
		this current will vanish after integrating $\textbf{k'}$.
		
		To conclude, we have proved the current density induced by impurity scattering is zero in both linear regime and nonlinear regime. Similar results can be found in \cite{PhysRevB.96.235134}\cite{PhysRevB.101.155204}. However, here we only prove that the term related to the dipole contributes nothing to the transport in nonlinear regime. We do not consider all the terms in general case.
		
		\section{More details of calculation on topological crystalline insulator}
		
		After constructing the effective Hamiltonian of strained single-layer
		graphene, we can calculate the eigenstates $H|u_{\textbf{k}}^{\pm}\rangle=\pm\epsilon_{\textbf{k}}|u_{\textbf{k}}^{\pm}\rangle$
		and Berry curvature of the model.
		
		\begin{equation}
			|u_{\textbf{k}}^{\pm}\rangle=\frac{1}{\sqrt{2}}\begin{pmatrix}\sqrt{1\pm\frac{\Delta}{2\epsilon_{\textbf{k}}}}\\
				\pm e^{i\theta}\sqrt{1\mp\frac{\Delta}{2\epsilon_{\textbf{k}}}}
			\end{pmatrix}
		\end{equation}
		where angle $\theta$ and $\epsilon_{\textbf{k}}$ is defined by $e^{i\theta}=\frac{A_{1}+iA_{2}}{\sqrt{A_{1}^{2}+A_2^{2}}}$ and $\epsilon_{\textbf{k}}=\sqrt{A_{1}^{2}+A_2^{2}+m^{2}}$.
		
		\begin{equation}
			\partial_{a}|u_{\textbf{k}}^{\pm}\rangle=\frac{1}{\sqrt{2}}\begin{pmatrix}\mp\frac{1}{2}\frac{1}{\sqrt{1\pm m/\epsilon_{k}}}\frac{\Delta}{2\epsilon_{\textbf{k}}^{2}}\frac{\partial\epsilon_{\textbf{k}}}{\partial k_{a}}\\
				\frac{1}{2}e^{i\theta}\frac{1}{\sqrt{1\mp m/\epsilon_{k}}}\frac{\Delta}{2\epsilon_{\textbf{k}}^{2}}\frac{\partial\epsilon_{\textbf{k}}}{\partial k_{a}}\pm\frac{i}{2}e^{i\theta}\frac{\partial\theta}{\partial k_{a}}\sqrt{1\mp\frac{\Delta}{2\epsilon_{\textbf{k}}}}
			\end{pmatrix}
		\end{equation}
		
		\[
		\Omega_{\textbf{k}z}^{\pm}=i(\langle\partial_{x}u_{\textbf{k}}^{\pm}|\partial_{y}u_{\textbf{k}}^{\pm}\rangle-\langle\partial_{y}u_{\textbf{k}}^{\pm}|\partial_{x}u_{\textbf{k}}^{\pm}\rangle)
		\]
		
		\begin{equation}
			=\mp\frac{\Delta}{4\epsilon_{\textbf{k}}^{3}}(\frac{\partial A_{1}}{\partial k_{x}}\frac{\partial A_{2}}{\partial k_{y}}-\frac{\partial A_{1}}{\partial k_{y}}\frac{\partial A_{2}}{\partial k_{x}})
		\end{equation}
		
		With the form of $A_{1}=v_{x}k_{x}$ and $A_{2}=v_{y}k_{y}$, you can obtain (28,29) automatically.
		
		For further step, we focus on the relaxation time.
		
		\begin{equation}
			\frac{1}{\tau_{\textbf{k}}^{m}}=\frac{2\pi }{\hbar}\int\frac{dk_{x}'dk_{y}'}{(2\pi)^{2}}\sum_{m,m'}\bra U_{\textbf{k}\textbf{k'}}^{mm'}U_{\textbf{k'}\textbf{k}}^{m'm}\ket\delta(\epsilon_{\textbf{k}}^{m}-\epsilon_{\textbf{k}'}^{m'})
		\end{equation}
		We here take $m=m'=+$ since not only there is no crossover near each valley but also we just basically assume $\mu>0$ for simplicity. In this way, only electrons from the conduction band contribute. So the relaxation time is:
		\begin{equation}
			\frac{1}{\tau_{\textbf{k}}^{+}}=\frac{2\pi }{\hbar}\int\frac{dk_{x}'dk_{y}'}{(2\pi)^{2}}\bra U_{\textbf{k}\textbf{k'}}^{++}U_{\textbf{k'}\textbf{k}}^{++}\ket\delta(\epsilon_{\textbf{k}}^{+}-\epsilon_{\textbf{k}'}^{+})
		\end{equation}
	With (C1), we have the Born approximation in the Bloch space\cite{PhysRevB.97.201301}:
	\begin{equation}
		 U_{\textbf{k}\textbf{k'}}^{++}=U\bra u_{k}^{+}|u_{k'}^{+}\ket=\frac{U_{0}}{2}[\sqrt{(1+\frac{\Delta}{2\epsilon_{\textbf{k}}})(1+\frac{\Delta}{2\epsilon_{\textbf{k'}}})}+e^{i(\theta'-\theta)}\sqrt{(1-\frac{\Delta}{2\epsilon_{\textbf{k}}})(1-\frac{\Delta}{2\epsilon_{\textbf{k'}}})}]
	\end{equation}
	\begin{equation}
		\bra U_{\textbf{k}\textbf{k'}}^{++}U_{\textbf{k'}\textbf{k}}^{++}\ket=\frac{n_{imp}U_{0}^{2}}{2}[1+\frac{\Delta^{2}}{4\epsilon_{\textbf{k}}\epsilon_{\textbf{k'}}}+cos(\theta'-\theta)\frac{\sqrt{v_x^{2}k_x^{2}+v_y^{2}k_y^{2}}\sqrt{v_x^{2}k_x'^{2}+v_y^{2}k_y'^{2}}}{\epsilon_{\textbf{k}}\epsilon_{\textbf{k'}}}]
	\end{equation}
	By neglecting the warping effect, we obtain the relaxation time:
	\begin{equation}
		\frac{1}{\tau_{\textbf{k}}^{+}}=\frac{n_{imp}U_{0}^{2}\mu}{4v_{x}v_{y}}(1+3\frac{\Delta^{2}}{4\mu^{2}})
	\end{equation}
		In SnTe, The energy bands are: $E=w_{y}k_{y}\pm\sqrt{v_{x}^{2}k_{x}^{2}+v_{y}^{2}k_{y}^{2}+(\frac{\Delta}{2})^{2}}$,which
		can be calculated analytically. Energy surface is given by:
		
		\begin{equation}
			\frac{(k_{y}+k_{0})^{2}}{s_{y}^{2}}+\frac{k_{x}^{2}}{s_{x}^{2}}=1
		\end{equation}
		
		In the formula: $s_{y}^{2}=(\frac{v_{y}^{2}}{v_{y}^{2}-w_{y}^{2}}E^{2}-(\frac{\Delta}{2})^{2})/(v_{y}^{2}-w_{y}^{2}),s_{x}^{2}=(\frac{v_{y}^{2}}{v_{y}^{2}-w_{y}^{2}}E^{2}-(\frac{\Delta}{2})^{2})/v_{x}^{2},k_{0}=\frac{w_{y}E}{v_{y}^{2}-w_{y}^{2}}$.
		With $k_{y}=-k_{0}+s_{y}sin\theta,k_{x}=s_{x}cos\theta$, 
		
		\begin{equation}
			dk_{x}dk_{y}=\begin{vmatrix}\frac{\partial k_{x}}{\partial E} & \frac{\partial k_{y}}{\partial E}\\
				\frac{\partial k_{x}}{\partial\theta} & \frac{\partial k_{y}}{\partial\theta}
			\end{vmatrix}dEd\theta=[\frac{v_{y}^{2}}{v_{x}(v_{y}^{2}-w_{y}^{2})^{3/2}}E-\frac{w_{y}}{v_{x}(v_{y}^{2}-w_{y}^{2})}\sqrt{\frac{v_{y}^{2}}{v_{y}^{2}-w_{y}^{2}}E^{2}-(\frac{\Delta}{2})^{2}}sin\theta]dEd\theta
		\end{equation}

		After preparation, let's focus on the conductivity and Berry curvature dipole. According to (19), we can derive $D_{y}$ first.
		
		\[
		D_{y}=\frac{1}{2}\int\frac{d^{2}k}{(2\pi)^{2}}\epsilon_{+}^{2}f_{0}(\epsilon_{\textbf{k}}^{+})\partial_{y}\Omega_{\textbf{k}z}^{+}
		\]
		
		\[
		=\frac{1}{2\hbar}\int_{E<\mu}\frac{dEd\theta}{(2\pi)^{2}}[\frac{v_{y}^{2}}{v_{x}(v_{y}^{2}-w_{y}^{2})^{3/2}}E-\frac{w_{y}}{v_{x}(v_{y}^{2}-w_{y}^{2})}\sqrt{\frac{v_{y}^{2}}{v_{x}(v_{y}^{2}-w_{y}^{2})}E^{2}-(\frac{\Delta}{2})^{2}}sin\theta](E-\mu)^{2}\frac{3\Delta v_{x}v_{y}^{3}(-k_{0}+s_{y}sin\theta)}{4(E+w_{y}k_{0}-w_{y}s_{y}sin\theta)^{5}}
		\]

		With calculation tools, we can give the result as Fig.1 in the context.
		Meanwhile, we can derive the conductivity with the relaxation time.
		\begin{equation}
			\chi_1=e\tau_{k_{F}}^{+}D_{y}=\frac{4 ev_{x}v_{y}}{n_{imp}U_{0}^{2}\mu(1+3\frac{\Delta^{2}}{4\mu^{2}})}D_{y}
		\end{equation}
	\[
	\chi_2=e\tau_{k_F}^{+}\int\frac{d^{d}k}{(2\pi)^{d}}\epsilon_{+}\frac{\partial\epsilon_{\textbf{k}}^+}{{\partial k_{y}}}f_{0}(\epsilon_{k}^{+})\Omega_{kz}^{+}
	\]
		\begin{equation}
		=\frac{4 ev_{x}v_{y}}{n_{imp}U_{0}^{2}\mu(1+3\frac{\Delta^{2}}{4\mu^{2}})}\int\frac{d^{d}k}{(2\pi)^{d}}\epsilon_{+}\frac{\partial\epsilon_{\textbf{k}}^+}{{\partial k_{y}}}f_{0}(\epsilon_{k}^{+})\Omega_{kz}^{+}
		\end{equation}
		We can numerically depict the result as Fig.2 above.
	\end{appendix}
	
\end{widetext}
\bibliography{BCD}
\end{document}